\documentclass{article}

\PassOptionsToPackage{numbers, compress}{natbib}

\usepackage[final]{neurips_2024_ml4ps}

\usepackage{amsmath}
\usepackage[utf8]{inputenc} 
\usepackage[T1]{fontenc}    
\usepackage{hyperref}       
\usepackage{url}            
\usepackage{booktabs}       
\usepackage{amsfonts}       
\usepackage{nicefrac}       
\usepackage{microtype}      
\usepackage{xcolor}         
\usepackage{graphicx}

\title{Convolutional Vision Transformer for Cosmology Parameter Inference}

\author{%
  Yash Gondhalekar\\
  \texttt{yashgondhalekar567@gmail.com} \\
  \And
  Kana Moriwaki\\
  Research Center for the Early Universe, Graduate School of Science, The University of Tokyo
}

\begin{document}

\maketitle

\begin{abstract}
Parameter inference is a crucial task in modern cosmology that requires accurate and fast computational methods to handle the high precision and volume of observational datasets. In this study, we explore a hybrid vision transformer, the Convolution vision Transformer (CvT), which combines the benefits of vision transformers (ViTs) and convolutional neural networks (CNNs). We use this approach to infer the $\Omega_m$ and $\sigma_8$ cosmological parameters from simulated dark matter and halo fields. Our experiments indicate that the constraints on $\Omega_m$ and $\sigma_8$ obtained using CvT are better than ViT and CNN, using either dark matter or halo fields. For CvT, pretraining on dark matter fields proves advantageous for improving constraints using halo fields compared to training a model from the beginning. However, ViT and CNN do not show these benefits. The CvT is more efficient than ViT since, despite having more parameters, it requires a training time similar to that of ViT and has similar inference times. The code is available at \url{https://github.com/Yash-10/cvt-cosmo-inference/}.
\end{abstract}

\section{Introduction}

An enthralling task in cosmology is accurately estimating the cosmological parameters describing the Universe from observational data, i.e., cosmological parameter inference. The widely accepted cosmological model, the $\Lambda$CDM ($\Lambda$ Cold Dark Matter), describes the Universe using a few parameters: $\Omega_m$ (the matter density, including normal and dark matter), $\Omega_{\Lambda}$ (the dark energy density; $\Lambda$, the cosmological constant, represents dark energy), $h$ (the Hubble parameter), $n_s$ (the spectral index of density perturbations), $\sigma_8$ (the variance in the matter distribution smoothed over spheres of radius 8 $h^{-1}$Mpc). The overwhelming amount of cosmological information from current and upcoming observational surveys \citep[e.g.,][]{Laureijs2011,Ivezic2019} will require sound statistical methodologies to achieve this goal.

Parameter inference aims to determine the posterior distributions of cosmological model parameters given a set of observations. Traditionally, this has been achieved by comparing the two-point correlation functions or power spectra of the tracers of large-scale structure (LSS) with theoretical predictions, or by using higher-order summary statistics that extract non-Gaussian information \cite[e.g.,][]{Cheng2020, Zurcher2021}. Such methods have analytically tractable likelihoods. However, predefined summary statistics inevitably fail to fully capture the rich non-Gaussian information at non-linear scales, which makes them suboptimal. Recently, `field-level inference' \cite{Cranmer2020} has gained a lot of attention as a potential alternative to these traditional techniques due to its ability to produce tighter constraints \cite[see, e.g.,][]{Leclercq2021,Andrews2023,deSanti2023,Lemos2023}. In this case, the likelihood is untractable since cosmological parameters are directly derived from the full, non-linear distribution of matter fields. Field-level inference allows access to higher-order information (e.g., from the phases of the fields), which is otherwise inaccessible through conventional summary statistics. Neural networks are a promising solution for field-level inference due to their demonstrated capabilities to extract features from complex data efficiently.

Since neural networks use the entire non-linear (and thus noisy) distribution of matter, they must effectively extract informative multi-scale features that help link those features to the underlying cosmological model parameters. Convolutional Neural Networks (CNNs) have consistently excelled in various tasks, primarily due to their localization through convolutional kernels, translation invariance property, and learning features hierarchically (i.e., local features in earlier layers and increasingly global features in later layers as its receptive field enlarges). Consequently, CNNs have become the dominant choice for cosmological parameter inference \cite[e.g.,][]{Ravanbakhsh2017,Ribli2019,Lazanu2021,VillaescusaNavarro2022}.

CNNs also have limitations because their receptive fields are constrained to grow larger as depth progresses, but that may not be necessary. The relatively newer Vision Transformers (ViTs) \cite{Dosovitskiy2020} do not take advantage of the strong inductive biases induced by convolutions in CNNs, allowing them to learn global spatial relationships through their self-attention layers even in the earlier layers of the model. ViTs break down an image into several patches, which are then flattened and considered tokens, similar to the terminology used in natural language processing. ViTs lack some biases, so they require large datasets for training, but given this constraint is satisfied, they have shown comparable or better performance than state-of-the-art CNNs such as ResNets. The applications of ViT for cosmological parameter inference are thus compelling, but only a few studies have explored their value \cite[][]{Huang2022,Hwang2023} and found them competitive with CNN. We hypothesize that combining the benefits of CNNs and ViTs may alleviate their individual deficiencies and improve parameter inference.

In this study, we perform likelihood-free inference to predict the marginal posterior mean and variance of the two cosmological parameters, $\Omega_m$ and $\sigma_8$, from simulated dark matter and halo distribution using the publicly available {\sc QUIJOTE} simulation suite as our dataset. We use a convolutional vision transformer (CvT) that combines the advantages of both CNNs and ViTs. CvT has been previously applied in \cite{Cao2024}, who found it better than CNN and ViT for classifying galaxy morphologies. 

\section{Method}

\paragraph{Data} We use the $N$-body simulations of the publicly available {\sc QUIJOTE} simulation suite \cite{Quijote2020} to obtain the DM density and halo catalogs; we use the friends-of-friends (FoF) halo catalogs. These simulations are performed with a box size of $L = 1\,h^{-1}$Gpc. We use standard Latin-hypercube simulations with massless neutrinos that contain $512^3$ cold dark matter particles and use outputs at $z = 0$. 
This simulation set contains 2000 simulation data, and we divide it into training, validation, and testing sets using an 80-10-10\% split. Since the splitting is performed at the simulation level, all data from a simulation are either in the training, validation, or testing set; such a non-random splitting is crucial to prevent obvious bias in the results \cite{Ribli2019-split}. All 1600 DM data are used for pretraining, but only 200 Halo data are used for finetuning.
All simulations have different random seeds with $\Omega_m$ varied in [0.1, 0.5], $\sigma_8$ varied in [0.6, 1.0], and the other astrophysical parameters ($\Omega_b$, $h$, $n_s$) varied within their appropriate ranges.

We project the particle and halo positions from the simulation snapshots onto a $256^3$ grid using the cloud-in-cell (CIC) scheme. All the halos detected in {\sc QUIJOTE} are contained in halo maps, i.e., we do not apply any mass cuts. Ten random two-dimensional maps (each of thickness $\sim$3.9$\,h^{-1}$Mpc) along each projection direction (X, Y, and Z) are selected, producing 30 maps from each simulation (these maps are individually used as inputs to the neural network). The overdensities are first calculated ($\rho / \overline{\rho}$), followed by a logarithm-like transformation given by $\log_{10} (1 + \rho / \overline{\rho})$ to reduce the dynamic range of the pixel values, followed by standardization using the mean and standard deviation of the transformed fields from the training set. Each cosmological parameter is individually normalized to $[0, 1]$ using the corresponding minimum and maximum values calculated from the training set.

\begin{figure}[htp]
    \centering
    \includegraphics[width=0.86\textwidth,keepaspectratio]{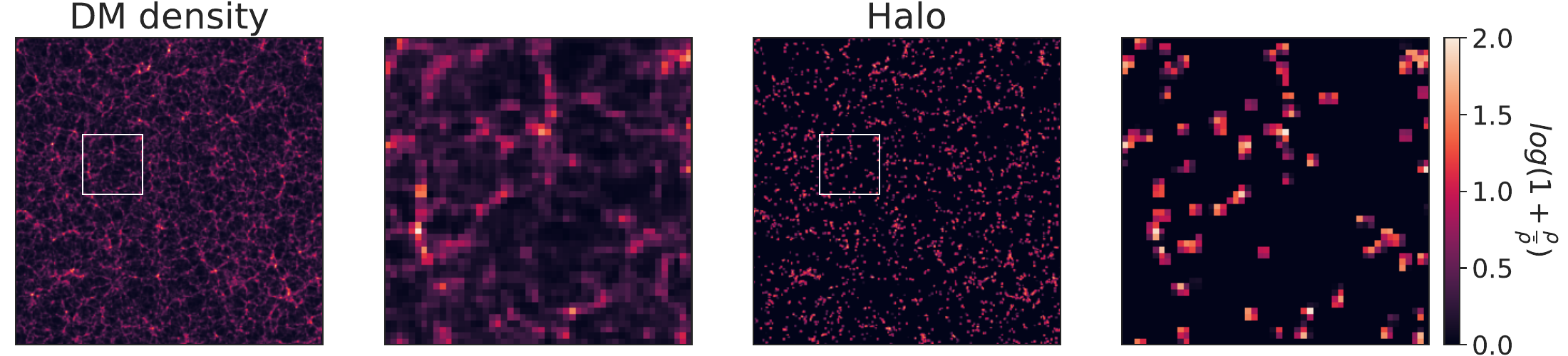}
    \caption{Sample visualization of the DM density and halo maps, extracted to show the same region from the simulation volume. The dimensions of the maps in the first and the third columns are $256 \times 256$ pixels, with thickness $\sim$$3.9\,h^{-1}$Mpc. The second and the fourth columns show a zoomed inset of $50 \times 50$ pixels (and the same thickness) to better illustrate the comparison between DM density and halo distribution.}
    \label{fig:dm-halo-visual}
\end{figure}

Fig.~\ref{fig:dm-halo-visual} shows an example comparison of the two-dimensional DM density and halo maps. It shows that the halos are biased tracers of the DM density as the former captures high local overdensities in the DM distribution, and thus leads to a visually sparser distribution.

\paragraph{Approach} We use a Convolutional vision Transformer (CvT) \citep{Wu2021}, which uses a multi-stage hierarchical structure (see Appendix~\ref{ref:cvt-fig} for visualization of the architecture), where each stage contains a convolutional token embedding layer followed by convolutional transformer blocks. The convolutional token embedding layers learn a convolution operation transforming input tokens into a new set of tokens. Its placement across different stages allows progressive spatial downsampling (i.e., reducing the number of tokens) together with increasing feature dimensions (i.e., increasing the width of tokens), and thus allows capturing local information as in CNNs. The convolutional transformer block uses a convolutional projection, implemented as a depth-wise separable convolution layer, instead of a linear projection used in traditional ViT. Ideologically, this transformer block generalizes the transformer in traditional ViT. Since local spatial relationships are modeled through the convolutional token embedding and projection, no positional encoding is required, which allows CvT to adapt to variable spatial resolution images. The use of efficient convolutions within the transformer in CvT also makes it computationally and memory-wise more efficient than traditional ViTs. The implementation is adapted from the \texttt{vit-pytorch} code\footnote{\url{https://github.com/lucidrains/vit-pytorch}}.
Specifically, we use the lightweight CvT-13 model, i.e., with a total of 13 transformer blocks containing 17.6M parameters and the default model hyperparameters as used in the original paper.

\paragraph{Training details} We modify the CvT architecture to perform a regression task to predict the two cosmological parameters, $\Omega_m$ and $\sigma_8$. Our model predicts the marginal posterior mean and variance for $\Omega_m$ and $\sigma_8$. The loss function is designed under the framework of moment networks \cite{Jeffrey2020} and used previously in works such as \cite{Villaescusa-Navarro2021} and \cite{VillaescusaNavarro2022}, and is given by:
\begin{equation}
    { \mathcal L } = \sum _{i=1}^{2}\mathrm{log}\left(\displaystyle \sum _{j\in \mathrm{batch}}{\left({\theta }_{i,j}-{\mu }_{i,j}\right)}^{2}\right)\\ + \sum _{i=1}^{2}\mathrm{log}\left(\displaystyle \sum _{j\in \mathrm{batch}}{\left({\left({\theta }_{i,j}-{\mu }_{i,j}\right)}^{2}-{\sigma }_{i,j}^{2}\right)}^{2}\right).
\end{equation}
where $\theta_{i,j}$ is the true value of parameter $i$ from simulation $j$ and $\mu_{i,j}$ and $\sigma_{i,j}$ are the network prediction of the mean and standard deviation of the marginal posterior of parameter $i$, respectively.
During training, the 2D maps are rotated randomly by 90, 180, or 270 degrees. A batch size of 16, Adam with decoupled weight decay optimizer (AdamW; \citealt{Loshchilov2017}) with weight decay of $10^{-5}$ and a learning rate of $5 \times 10^{-6}$ is used. The learning rate is reduced by a factor of 0.3 if the validation loss does not improve for five epochs. Dropout is not used during training. Training is performed for 30 epochs in all experiments, and the model weights corresponding to the lowest validation loss are used for inference. Weights \& Biases \citep{Wandb2020} was used to track training and evaluation. No specific hyperparameter tuning has been performed.

\section{Results}

\paragraph{Experimental details and evaluation} We pretrain CvT on DM fields and report the test results. We also show test results when the pretrained model is finetuned on halo fields and compare its performance against the model trained from scratch on halo fields. Before finetuning, we only reinitialized the fully connected MLP head. We chose not to freeze any weights, as doing so only provided marginally better constraints on $\Omega_m$: the RMSE was almost the same, with error bars increasing by about 0.014, reflecting the RMSE better, but constraints on $\sigma_8$ were significantly worse, with the RMSE increasing by about 0.014 and error bars underpredicted by about 0.03, than not freezing. The optimal pretrained model on DM data was found after 28k iterations, the optimal finetuned model on halo data after 2.4k iterations, and the optimal model trained from scratch on halo data after 1.4k iterations, although it had a higher validation loss than using transfer learning.

We used two metrics to evaluate the prediction for each cosmological parameter: the root mean squared error (RMSE), defined as $\mathrm{RMSE}_i = \sqrt{\frac{\sum_{j=1}^{N}(\theta_{i,j} - \mu_{i,j})^2}{N}}$, where $N$ is the number of test examples, and $\bar{\sigma_i} = \frac{1}{N} \sum_{j=1}^{N}\sigma_{i,j}$ denotes the averaged errors. $\mathrm{RMSE}_{i}$ determines the accuracy of the predictions, and $\sigma_i$ denotes the 1$\sigma$ error in the prediction of the parameter value.

\begin{figure}[htp]
    \centering
    \includegraphics[width=.285\textwidth,keepaspectratio]{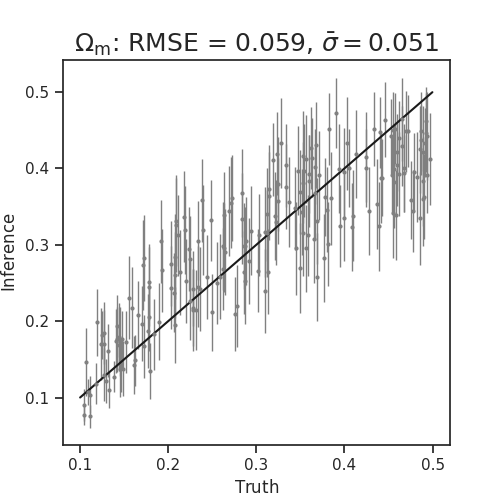}
    \includegraphics[width=.285\textwidth,keepaspectratio]{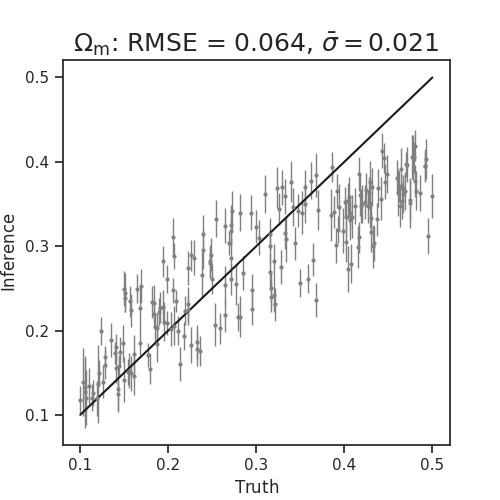}
    \includegraphics[width=.285\textwidth,keepaspectratio]{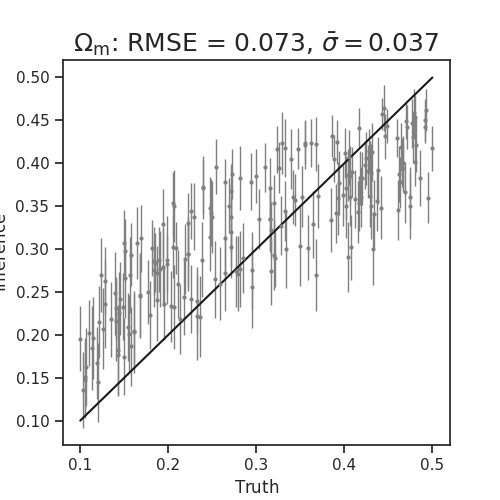}
    \includegraphics[width=.285\textwidth,keepaspectratio]{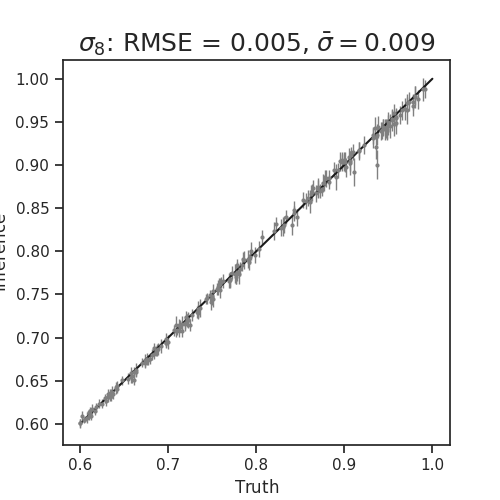}
    \includegraphics[width=.285\textwidth,keepaspectratio]{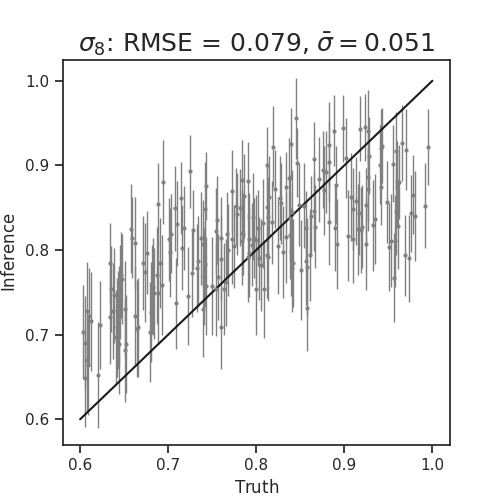}
    \includegraphics[width=.285\textwidth,keepaspectratio]{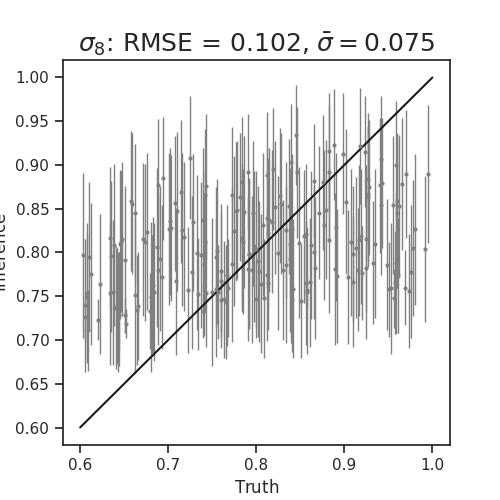}
    \caption{Comparison of predicted (y-axis) and ground-truth (x-axis) $\Omega_m$ and $\sigma_8$ cosmological parameters on the test set. The first, second, and third columns show the test results of pretraining on DM data, transfer learning on halo data, and training a model from scratch on halo data, respectively. Each data point shows the averaged values and errors across all 30 2D maps from a single 3D simulation volume. The title of each panel shows the RMSE and $\bar{\sigma}$ (see text for description).}
    \label{fig:param-preds}
\end{figure}

\paragraph{Prediction performance} Fig.~\ref{fig:param-preds} shows the predictions versus the true parameter values for DM pretraining (a), halo transfer learning (b), and halo training from scratch (c), from left to right. $\sigma_8$ predictions for (a) show excellent agreement with a near 1:1 relationship (RMSE = 0.005 and appropriately small error bars), while $\Omega_m$ predictions are moderately good (RMSE = 0.059) and appropriate error bars. To put these values in context, we report the RMSE in the case of a constant prediction equal to the mean of the true value and obtain RMSE = 0.118 for both parameters.

Using the pretrained model from dark matter and transfer learning using halo fields (case b) gives slightly worse constraints for $\Omega_m$ than (a) (RMSE = 0.064 and underestimated error bars), but the $\sigma_8$ constraints are more prominently deteriorated (RMSE = 0.079 and slightly underestimated error bars). This deterioration is not unexpected, as halos are biased tracers of the underlying dark matter field, so they contain less pertinent information. It is currently elusive why the error bars for $\Omega_m$ are severely underestimated, but the fact that this also happens for (c) suggests that this is not necessarily due to transfer learning but a characteristic feature when using halo fields. The error bars for $\sigma_8$ are also underestimated, but this is less severe than $\Omega_m$.

For (b), it can be seen that large $\Omega_m$ values tend to be underestimated, small $\sigma_8$ values are overestimated, whereas large $\sigma_8$ values are underestimated. So, predictions near the edges are affected and there is a tendency to regress towards the mean of the cosmological parameter set\footnote{Although we intend to talk about the mean of the `test' parameter set here, we have checked the mean of the training parameter set is also similar which is because we randomly split the simulations.}. The RMSE for constant prediction is 0.121 and 0.109 for $\Omega_m$ and $\sigma_8$, respectively. Thus, this accuracy is still better than simply predicting the mean value. We do not have a clear explanation for these biases, but they may be due to insufficient expressiveness of the MLP head (since we only use a single-layered MLP) or due to overfitting (see \cite{Pan2020} and \cite{Ntampaka2020}, respectively).

The constraints in (c) are worse than in (b) as shown by the lower values of {\sc RMSE} and $\bar{\sigma}$, and therefore the transfer learning approach (DM pretraining followed by halo finetuning) seems more beneficial than training on halo data from scratch. This can be expected because the large-scale features in the DM and halo fields are similar, so the pretrained weights of the model that is trained on DM data serve as a better starting point to learn features from the halo fields.

\paragraph{Comparison with traditional ViT} We compare the CvT (used in this work) with the simpler version of the traditional ViT architecture discussed in \citet{Beyer2022}, which we dub the `ViT'\footnote{Note that this is a slightly modified version of the original ViT architecture proposed in \cite{Dosovitskiy2020}}. We used a patch size of 8 for the ViT, but other common hyperparameters are the same as CvT. The results for (a), (b), and (c) for $\Omega_m$ are as follows: RMSE = 0.066 and $\bar{\sigma}$ = 0.254, RMSE = 0.068 and $\bar{\sigma}$ = 0.299, RMSE = 0.074 and $\bar{\sigma}$ = 0.281. Thus, for (a) and (b), ViT is less accurate than CvT. For (c), the RMSEs are similar, but the error bar for CvT is more representative of the accuracy. For $\sigma_8$, the results for (a), (b), and (c) are: RMSE = 0.1 and $\bar{\sigma}$ = 0.24, RMSE = 0.106 and $\bar{\sigma}$ = 0.314, RMSE = 0.112 and $\bar{\sigma}$ = 0.247. However, the predictions are `near-flat'\footnote{The predicted parameters are visually similar irrespective of the true parameter value when visualized like Fig.~\ref{fig:param-preds}.} in all cases. Thus, CvT can constrain the cosmological parameters more tightly than ViT, especially $\sigma_8$.

\paragraph{Comparison with CNN} The CNN architecture consists of five convolutional layers and batch normalization, followed by a fully connected layer that predicts the mean and standard deviation, just like the ViT, and other common hyperparameters are the same as CvT. The results for (a), (b), and (c) for $\Omega_m$ are as follows: RMSE = 0.073 and $\bar{\sigma}$ = 0.086, RMSE = 0.21 and $\bar{\sigma}$ = 0, RMSE = 0.106 and $\bar{\sigma}$ = 0.139. CNN is less accurate than ViT and CvT, and also yields an overconfident prediction for (b) ($\bar{\sigma}$ = 0). For $\sigma_8$, the results for (a), (b), and (c) are: RMSE = 0.035 and $\bar{\sigma}$ = 0.075, RMSE = 0.151 and $\bar{\sigma}$ = 0, RMSE = 0.116 and $\bar{\sigma}$ = 0.137. CNN is better than ViT to predict $\sigma_8$ for (a), but is worse than both ViT and CvT for all other cases. 

\paragraph{Execution time} The ViT used in this work contains far fewer parameters (1.6M vs. 17.6M) but requires only marginally shorter training time than the CvT ($\sim$2.6 vs. 2.7 hours) for the DM pretraining. Thus, the operations in CvT are more efficient than those in ViT, probably due to the introduction of convolutions and the convolutional projection operation \citep{Wu2021}. At test time, CvT requires $\sim$24 seconds, whereas ViT requires $\sim$19 seconds for inference on 6000 maps (this experiment was performed when the model was trained using halo data from scratch). Although CvT is cumulatively slightly slower, the per-map inference times are almost the same.

\section{Conclusion}

We have applied the convolution-based vision transformer (CvT) to infer the $\Omega_m$ and $\sigma_8$ cosmological parameters using data obtained from the {\sc QUIJOTE} simulation. We find that CvT constrains both parameters better than CNN and ViT when using dark matter and halo fields. CNN is found to be more beneficial than ViT only for inferring $\sigma_8$ from dark matter fields, whereas ViT outperforms in all other cases. Pretraining CvT on dark matter fields has proven beneficial in better constraining the parameters when finetuned on halo fields rather than training a model from scratch on halo data, but these benefits are not apparent for CNN and ViT. One possible interpretation is that CvT is able to effectively leverage the large-scale structure similarities between dark matter and halo fields; however, more detailed tests are necessary to validate this finding. The demonstrated constraining power of CvT is noteworthy given that it was finetuned using 8$\times$ lesser data than pretraining. As a result, it may be advantageous to apply CvT on data such as galaxy distribution, which require hydrodynamic simulations that are often computationally prohibitive. We also briefly demonstrate that CvT is more efficient than ViT due to the use of convolutions and has a similar inference time to it.

Some future aims of this work are to interpret CvT, apply it to real data and develop guidelines for observational cosmologists instructing the regions to look at in the data, and integrate it with data simulation approaches based on deep learning (i.e., emulators).

\begin{ack}
This work was supported by JSPS KAKENHI Grant Number 23K03446.
\end{ack}


\bibliographystyle{plainnat}
{\footnotesize \bibliography{neurips_bib}}














\appendix

\section{Architecture of the convolutional vision transformer}\label{ref:cvt-fig}

\begin{figure}[htp]
    \centering
    \includegraphics[width=.9\textwidth,keepaspectratio]{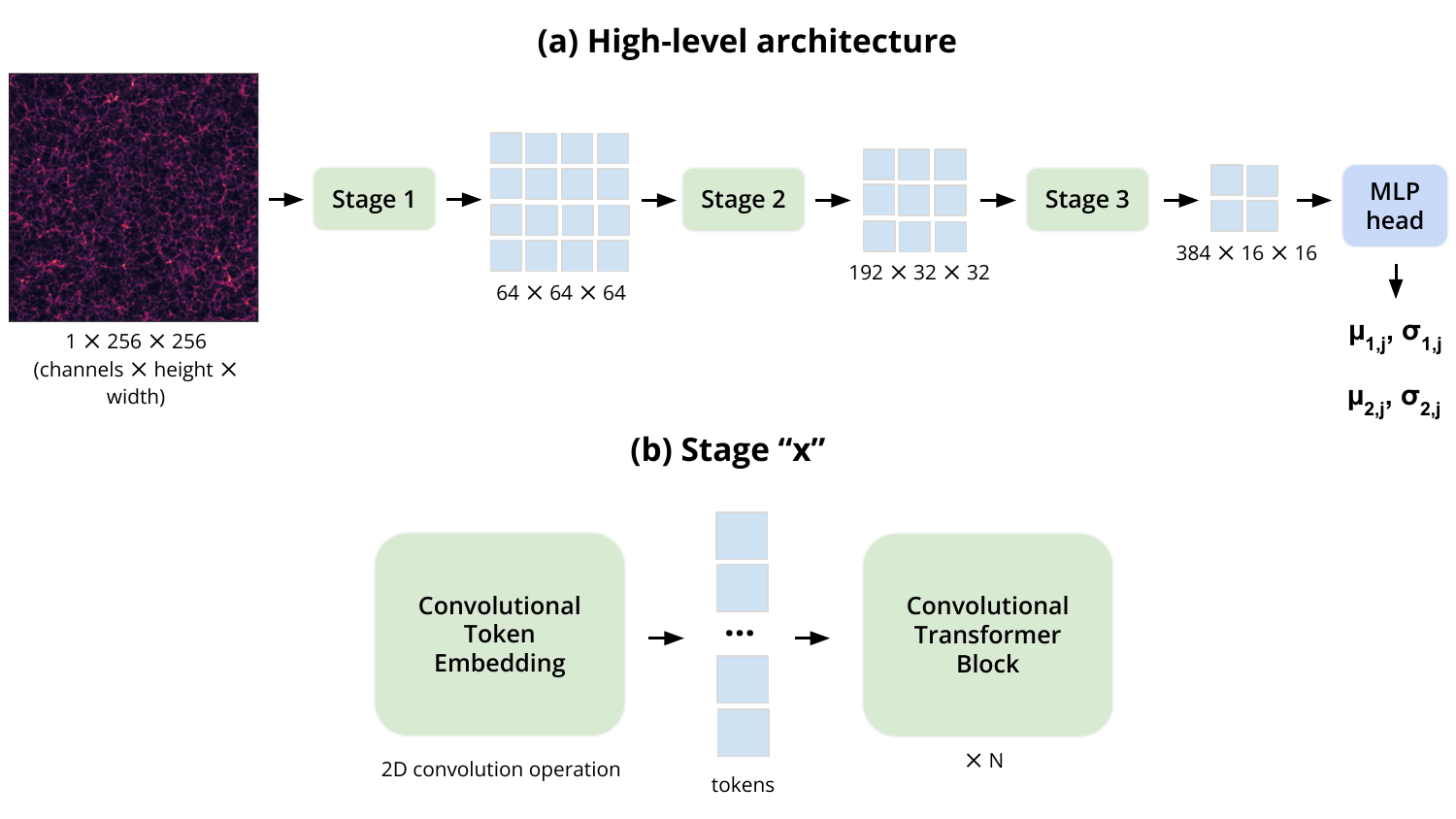}
    \includegraphics[width=.9\textwidth,keepaspectratio]{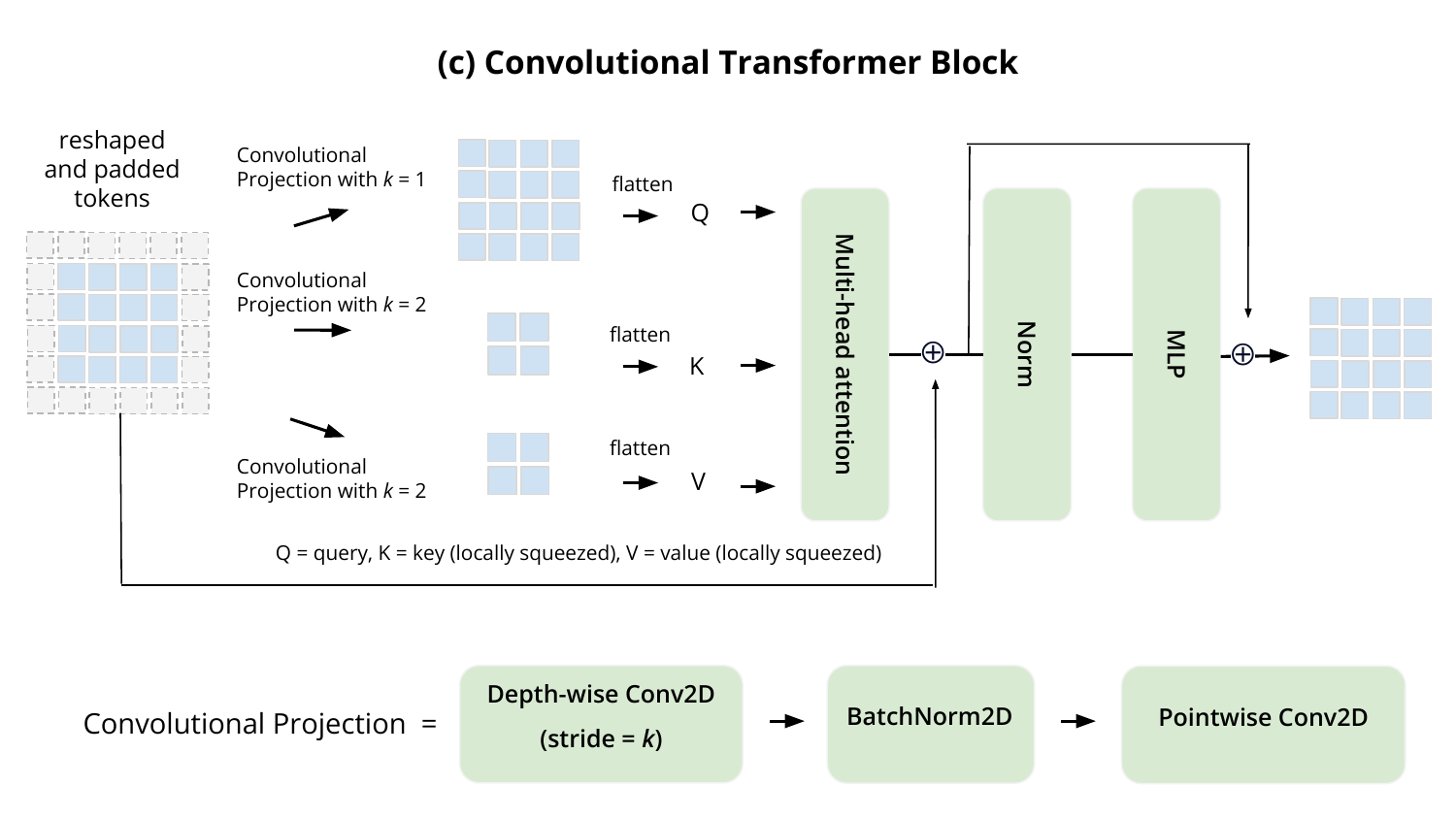}
    \caption{Architecture of the CvT network, demonstrating how convolutions and vision transformers are integrated in CvT. (a) shows CvT's hierarchical multi-stage pipeline, allowing spatial downsampling while increasing the no. of feature maps. The MLP head performs the regression to output the mean and standard deviation of the marginal posteriors of the two cosmological parameters (see ``Training details'' for notation). (b) shows each stage's pipeline, consisting of a convolutional token embedding layer followed by $N$ convolutional transformer blocks. (c) details the architecture of the convolutional transformer block, which contains convolutional projection to project the query, key, and values as the first step, which is consequently passed to the multi-head self-attention module, and then normalization layer and MLP. No regression token is used.}
    \label{fig:cvt-fig}
\end{figure}

Fig.~\ref{fig:cvt-fig} shows the architecture of the CvT network. The primary modifications in CvT compared to the traditional ViT are the presence of a convolutional token embedding layer, whose presence across multiple stages resembles the design of CNNs, and the presence of a convolutional projection instead of a linear projection.

\end{document}